\documentclass[
  cha,
  aip,
  reprint,
  showpacs
]{revtex4-1}

\usepackage[T1]{fontenc}
\usepackage[utf8]{inputenc}
\usepackage{amsmath}
\usepackage{amssymb}
\usepackage{graphicx}
\usepackage{grffile}
\usepackage{refstyle}
\usepackage{txfonts}
\usepackage[normalem]{ulem}
\usepackage[
  citecolor=blue,
  colorlinks,
  linkcolor=blue,
  urlcolor=blue,
]{hyperref}

\begin{document}
\title{Synchronizing noisy nonidentical oscillators by transient uncoupling}

\author{Aditya Tandon}
\email{adityat@iitk.ac.in}
\affiliation{
  Department of Physics,
  Indian Institute of Technology Kanpur,
  Uttar Pradesh 208016, India
}
\author{Malte Schr\"{o}der}
\email{malte@nld.ds.mpg.de}
\affiliation{Network Dynamics, Max Planck Institute for Dynamics and Self-Organization (MPIDS), 37077 G\"ottingen, Germany}
\author{Manu Mannattil}
\email{mmanu@iitk.ac.in}
\affiliation{
  Department of Physics,
  Indian Institute of Technology Kanpur,
  Uttar Pradesh 208016, India
}
\author{Marc Timme}
\email{timme@nld.ds.mpg.de}
\affiliation{Network Dynamics, Max Planck Institute for Dynamics and Self-Organization (MPIDS), 37077 G\"ottingen, Germany}
\affiliation{
  Department of Physics,
  Technical University of Darmstadt,
  64289 Darmstadt, Germany
}
\author{Sagar Chakraborty}
\email{sagarc@iitk.ac.in}
\affiliation{
  Department of Physics,
  Indian Institute of Technology Kanpur,
  Uttar Pradesh 208016, India
}
\affiliation{
  Mechanics and Applied Mathematics Group,
  Indian Institute of Technology Kanpur,
  Uttar Pradesh 208016, India
}

\begin{abstract}
Synchronization is the process of achieving identical dynamics among coupled identical units. If the units are different from each other, their dynamics cannot become identical; yet, after transients there may emerge a functional relationship between them---a phenomenon termed `generalized synchronization'. Here we show that the concept of transient uncoupling, recently introduced for synchronizing identical units, also supports generalized synchronization among nonidentical chaotic units. Generalized synchronization can be achieved by transient uncoupling even when it is impossible by regular coupling. We furthermore demonstrate that transient uncoupling stabilizes synchronization in the presence of common noise. Transient uncoupling works best if the units stay uncoupled whenever the driven orbit visits regions that are locally diverging in its phase space. Thus, to select a favorable uncoupling region, we propose an intuitive method that measures the local divergence at the phase points of the driven unit's trajectory by linearizing the flow and subsequently suppresses the divergence by uncoupling.
\end{abstract}
\pacs{05.45.Xt, 05.45.Gg}
\maketitle

\begin{quotation}
About two decades ago, researchers discovered that two coupled identical chaotic systems may synchronize and achieve identical dynamics.
Earlier, this seemed counterintuitive, because chaotic systems exhibit sensitive dependence on initial conditions.
This concept of synchronization was further generalized to include coupled nonidentical chaotic oscillators.
Recently, another seemingly surprising result highlighted how occasional uncoupling of two identical chaotic units during their simultaneous time evolution induces synchronization.
This phenomenon was named transient uncoupling.\cite{sch2015} Here, we demonstrate how transient uncoupling effects generalized synchronization between nonidentical chaotic systems.
Transient uncoupling may also suppress noise induced desynchronization in such coupled systems.
Additionally, we explain why the counterintuitive effects of transient uncoupling are actually not unexpected.
\end{quotation}

\section{Introduction}
Synchronization, the coordination of individual systems to achieve identical dynamics, is a ubiquitous phenomenon realized in coupled dynamical systems. The study of synchronization goes back to the seventeenth century when Christiaan Huygens reported the synchronization of two pendulum clocks suspended from a single horizontal beam. Since then synchronization and related phenomena have been found in diverse contexts---biological and ecological~(fireflies, cricket chirps, networks of neurons, pacemaker cells, etc.);\cite{vic1995, bla1999,lew2002, agu2011} physical and engineering;\cite{tim2006,stw2010,tan2014,mea2014} and sociological~(crowd clapping together, crowd marching, etc.).\cite{str2005} In the last two decades, the theories on synchronization in chaotic systems has attracted a lot of attention.\cite{prk2003,bjps2009}

The phenomenon of synchronization has been systematically classified into many types, viz.,~complete or identical synchronization,\cite{yam1983,pc1990} phase synchronization,\cite{rpk1996,roh1998} imperfect phase synchronization, burst synchronization,\cite{izh2000,djd2004} lag synchronization,\cite{rpk1997} intermittent lag synchronization,\cite{rpk1997,boc2000} generalized synchronization (either weak or strong) \cite{rul1995,koc1996,hoy1997,pyr1996} and so on. Researchers have tried to invent a unified definition to encompass these different types, and also have contemplated extending the definition to infinite dimensional systems described by partial differential equations and/or systems where noise is present.\cite{bk2000}

One may recall that two identical subsystems are said to be in complete synchrony when, irrespective of the initial conditions, their variables are exactly identical in the limit:~$ \mbox{time } t \to \infty$. We then say that the coupled `system' is in a stable synchronized state. {However, given a collection of subsystems, it is a priori not at all obvious which type of synchronization may result for different types of coupling.} Moreover, a pair of chaotic subsystems are typically only synchronizable for a particular range of parameter values.
Therefore a practical and general methodology that induces synchronization among coupled subsystems is desirable. The recently proposed method of transient uncoupling \cite{sch2015} is one such interesting method. It has been reported to induce complete synchronization in a pair of identical chaotic subsystems that are otherwise unsynchronized for the same coupling strength.

In more commonly found pairs of nonidentical subsystems complete synchronization cannot take place, but generalized synchronization, i.e., functional dependence between the asymptotic states of the subsystems, can potentially be observed: Consider two identical or nonidentical chaotic subsystems,~$X$ and~$Y$, described by
\begin{equation} \label{eq:FG}
  \dot{\mathbf{x}} = \mathbf{F}(\mathbf{x})\,\,\mathrm{and}\,\, \dot {\mathbf{y}} = \mathbf{G}(\mathbf{y})
\end{equation}
 respectively, where~$\mathbf{x}(t),\mathbf{y}(t) \in~\mathbb{R}^N$ are the states of the two subsystems at time $t$. Let subsystem~$Y$ be unidirectionally and diffusively coupled to subsystem~$X$ as follows:
\begin{equation} \label{eq:Couple}
  \begin{aligned}
\dot{\mathbf{y}} = \mathbf{G}(\mathbf{y}) + \varepsilon C \cdot(\mathbf{x-y}),
  \end{aligned}
\end{equation}
where~$C \in \mathbb{R}^{N \times N}$ is a constant, time-independent, coupling matrix and~$\varepsilon$ is the coupling strength parameter.\cite{pea1997} A generalized synchronization state is said to exist when there is a map~$\mathbf{\Phi}$ taking the trajectories from the driver~$X$ onto the trajectories of the driven unit~$Y$, i.e.~$\mathbf{y}(t) = \mathbf{\Phi(x}(t))$.
In this article, we ask whether transient uncoupling supports and may actually induce generalized synchronization among nonidentical chaotic oscillators.

{Furthermore, natural systems are not only usually nonidentical but are also invariably affected by noise.} The role noise plays in synchronization has been a source of debate.\cite{her1995,lai1998,rea2001,wcw2000,zea2002,zho2002,dfh2004,kea2006,hkm2006,gll2006} Studying synchronization in settings where the systems are additionally driven by the same random forcing is important in the context of neuroscience\cite{izh2000,mai1995} and ecology.\cite{hsc2010} It has been observed that common noise when supplied to nonidentical systems, either enhances or destroys generalized synchronization depending on the system details.\cite{gll2006} We show here that {transient uncoupling stabilizes the synchronized states even in the presence of common noise.}
\section{Transient uncoupling and generalized synchronization}
  Transient uncoupling\cite{sch2015} is defined as the switching off of diffusive coupling between two identical subsystems (i.e., when $\mathbf{F=G}$ in Eq.~(\ref{eq:FG})) when the phase trajectory enters a particular subset of the driven subsystem. In other words, we multiply the coupling strength parameter in Eq.~(\ref{eq:Couple}) by a factor~$\overline{\chi}(\mathbf{y})$ given by
  \begin{equation} \label{eq:TUS}
    \begin{aligned}
      \overline{\chi}(\mathbf{y}) =
        \left\{
          \begin{array}{ll}
            0  & \mbox{for } \mathbf{y} \in \mathbb{U} \\
            1 & \mbox{for } \mathbf{y} \notin \mathbb{U},
          \end{array}
        \right.
    \end{aligned}
  \end{equation}
  where~$\mathbb{U} \subseteq \mathbb{R}^N$ (the phase space of the driven unit). In principle, for identical subsystems, defining $\mathbb{U}$ as a subset of the driver unit also works.
  The new dynamics of the driven unit~$Y$ is described by
    \begin{equation} \label{eq:UnCouple}
      \begin{aligned}
        \dot{\mathbf{y}} = \mathbf{G(y) + \varepsilon } \overline{\chi}\mathbf{(y)} C \mathbf{\cdot(x-y),}
      \end{aligned}
    \end{equation}
  whereas the dynamics of the driver unit~$X$ remains unchanged. Transient uncoupling has been shown to completely synchronize identical subsystems in a far wider range of coupling strengths. This paper looks to establish the possible effectiveness of this scheme on nonidentical subsystems. However for nonidentical subsystems aiming for generalized synchronization is more appropriate.

  With a view to quantitatively characterize generalized synchronization, although there are other methods,\cite{pc1990,mos2012} we have chosen to work with arguably the easiest one: {the auxiliary system method}.\cite{aba1996} For the convenience of the readers, we briefly summarize it below. If a generalized synchronization state exists between~$X$ and~$Y$,~$\mathbf{y}(t)$ is functionally determined by~$\mathbf{x}(t)$ in the asymptotic limit. We consider an exact replica of~$Y$,~$Y'$ (say), that is identically coupled to $X$:
  \begin{equation} \label{eq:Replica}
  \begin{aligned}
    \dot{\mathbf{y'}}= \mathbf{G(y')} + \varepsilon C\cdot(\mathbf{x}-\mathbf{y'}).
  \end{aligned}
  \end{equation}
   Now, the crucial argument is that in order to argue for generalized synchrony between $X$ and $Y$ (i.e., existence of $\mathbf{\Phi}$), one has to establish that $Y$ and $Y'$ are in complete synchrony.\cite{pyr1996} It is known that the linear stability of the manifold~$\mathbf{y}'(t)=\mathbf{y}(t)$ is equivalent to the linear stability of the manifold of the generalized synchronized motions in the total phase space $X\oplus Y$.\cite{aba1996} Identical synchronization in~$Y\oplus Y'$ is quantified by the maximal Lyapunov exponent of the transverse system~$\mathbf{y_\perp =: y - y'}$ described by
  \begin{eqnarray}
      \dot{\mathbf{ y}}_\perp &=&\mathbf{G(y) - G(y') -\varepsilon} C \cdot \mathbf{(y - y')}\,,\nonumber \\
      &\approx& [\mathbf{J}(\textbf{y}_s(t)) - \mathbf{\varepsilon} C]\mathbf{y_{\perp}},\label{eq:Transverse}      \end{eqnarray}
where~$\mathbf{J(y)}$ is the matrix~$\partial\mathbf{G}/\partial{\mathbf{y}}$ and~$\mathbf{y}_s(t)$ is the synchronous state. For the state~$\mathbf{y}_s(t)$ to be stable, the maximal transverse Lyapunov exponent
  \begin{equation} \label{eq:lyap}
    \begin{aligned}
      \lambda_{\textrm{max}}^{\perp} = \lim_{t \to \infty} \frac{1}{t} \ln \frac{\mathbf{\|y_\perp(}t)\|}{\mathbf{\|y_\perp(}0)\|}
    \end{aligned}
  \end{equation}
  must be negative.\cite{prk2003}

In order to fully appreciate our results, we analytically discuss generalized synchronization in two coupled identical logistic maps:
\begin{subequations}
\begin{eqnarray}
&&x_{n+1}=Q(x_n)\,,\\
&&y_{n+1}=Q(y_n)+\varepsilon[Q(x_n)-Q(y_n)]\,.
\end{eqnarray}
\end{subequations}
Here, $Q(x)=4x(1-x)$. The auxiliary map, thus, is
\begin{eqnarray}
y'_{n+1}=Q(y'_n)+\varepsilon[Q(x_n)-Q(y'_n)]\,.
\end{eqnarray}
If a relation of the form $y_n=\mathbf{\Phi}(x_n)$ exists asymptotically, then the two logistic subsystems are definitionally in synchrony. Evidently, here $\mathbf{\Phi}$ is a scalar function. One classifies this synchronization as strong or weak depending on whether $\mathbf{\Phi}$ is differentiable or not.
It is easy to show\cite{pyr1996} that as $\varepsilon$ is increased weak synchronization (nondifferentiable $\mathbf{\Phi}$) precedes strong synchronization (differentiable $\mathbf{\Phi}$): first there is synchronization between $y$ and $y'$, then $x$ and $y$ synchronize at a higher coupling strength. Now, let us weigh the effect of transient uncoupling. Since the driven and the driver subsystems are identical, we define $\mathbb{U}$ as the subset $[c_1,c_2]\subseteq[0,1]$ of the phase space of the {driver}. Here, $c_1,c_2\in[0,1]$ and $c_1\le c_2$. Using the invariant probability density $\rho_\textrm{inv}(x)=1/[\pi\sqrt{x(1-x)}]$ of the logistic map, it is straightforward to calculate the transverse Lyapunov exponent~($\lambda_{yy'}^\bot$, conditioned on $x_n$) of the invariant manifold $y=y'$ and also the transverse Lyapunov exponent~($\lambda_{xy}^\bot$) of the invariant manifold $x=y$. These are respectively given by
\begin{eqnarray}
&&\lambda_{yy'}^\bot=\theta\ln(1-\varepsilon)+\lim_{n\rightarrow\infty}\frac{1}{n}\sum_{i=1}^{n}\ln|Q'(y_n)|\,,\,\,\textrm{and}\\
&&\lambda_{xy}^\bot=\theta\ln(1-\varepsilon)+\ln2\,,
\end{eqnarray}
where $\theta=:\int_{0}^{c_1}\rho_\textrm{inv}(x)dx+\int_{c_2}^{1}\rho_\textrm{inv}(x) dx\le1$. For weak synchronization only $\lambda_{yy'}^\bot<0$, whereas for strong synchronization $\lambda_{xy}^\bot<0$ as well. The latter's threshold coupling parameter is thus given by $\varepsilon=1-1/2^{1/\theta}$ at which $\lambda_{xy}^\bot=0$. Hence, one notes that as $\theta$ decreases (meaning more uncoupling), $\varepsilon$ increases. In other words, the subsystems can now synchronize only at higher values of $\varepsilon$. This seems very natural as one would expect uncoupling to disrupt synchronization. A similar conclusion holds for weak synchronization as well. An immediate question then would be: \emph{How and why, if at all, should transient uncoupling be effective in inducing generalized synchronization?}
\section{Transient uncoupling in coupled nonidentical oscillators}
Let us conduct numerical experiments on the effect of transient uncoupling on systems that are known to exhibit generalized synchronization. To this end, we choose a system consisting of the Lorenz subsystem\cite{lor1963} and the dynamo subsystem.\cite{mai1999} For this system, noise has been shown to improve or destroy the stability of the generalized synchronization state depending on the direction of coupling.\cite{gll2006} This interplay between noise and generalized synchronization has been exploited later in this paper and this is our primary motivation for choosing this particular synchronizable system. In principle, our study on this system is quite general and could be done using any two coupled nonidentical subsystems.

If such nonidentical oscillators are coupled continuously at all times (no uncoupling), they exhibit generalized synchronization.\cite{gll2006} This means that, by definition, there is a functional relationship between the driver and the driven variables for the orbits on the overall chaotic attractor. Such a function, however, is neither known a priori nor is it easy to find it a posteriori. Thus, in order to understand the action of transient uncoupling on the synchrony of coupled subsystems, we prepare two nonidentical subsystems such that the form of the function relating the drive and the driven variables are known beforehand. In what follows, we start with such a system and subsequently explore the other one, viz., the dynamo--Lorenz system.
\subsection{The R\"ossler and transformed R\"ossler system}
Refer back to Eqs.~(\ref{eq:FG}) and (\ref{eq:Couple}), and let $\mathbf{F}=\mathbf{G}$. Consider the identical oscillators to be coupled R\"ossler oscillators defined by
$\mathbf{F}(\mathbf{x}) = \left(-(x_2 + x_3), x_1 + ax_2, b + x_3(x_1 - c)\right)^\mathsf{T}$ \cite{roessler76}
and $C \in \mathbb{R}^{3 \times 3}$, where $C_{ij} = 1$ for $i=j=1$ and $C_{ij} = 0$ otherwise. In addition, we choose $a = b = 0.2$, $c = 5.7$, and we notationally define $\mathbf{x}=:(x_1,x_2,x_3)^\mathsf{T}$. It has been comprehensively shown\cite{sch2015} that transient uncoupling remarkably enhances the range of coupling parameter $\varepsilon$ for which the aforementioned system is synchronized.

Let us perform a nonlinear transformation of driven variables:\cite{rul1995}
\begin{equation}
z_1=y_1\,,\,z_2=y_2+0.4y_3-0.008y_3^2\,,\,z_3=y_3\,,\label{eq:tn}
\end{equation}
and consequently the driven subsystem explicitly becomes
\begin{subequations}
\begin{eqnarray}
&&\dot{z}_1=-[z_2+0.6z_3+0.008z_3^2]-\varepsilon(z_1-x_1)\,,\\
&&\dot{z}_2=z_1+a(z_2-0.4z_3+0.008z_3^2)+\,\nonumber\\
&&\phantom{z_2=}+(0.4-0.016z_3)[b+z_3(z_1-c)]\,,\\
&&\dot{z}_3=b+z_3(z_1-c)\,,
\end{eqnarray}
\end{subequations}
which we call the transformed R\"ossler equations.
If the coupled R\"ossler oscillators are synchronized then so should the driver R\"ossler subsystem and the driven transformed R\"ossler equations. Technically speaking, these two subsystems, being nonidentical, cannot exhibit complete synchronization. However, what is important to note is that, by definition, they should be in generalized synchrony whenever the coupled identical R\"ossler subsystems are in complete synchrony; and therefore, because transient uncoupling is effective on the latter, it is expected to be effective in inducing synchrony in the former system as well.

However, before embarking on the numerical results, we highlight a caveat worth paying attention to. When the diffusive coupling parameter of the two identical R\"ossler subsystems is varied from zero to infinity, complete synchronization is seen for a small range of values of coupling parameter: $\varepsilon\in(\varepsilon_{c1},\varepsilon_{c2})$ where $0<\varepsilon_{c1}<\varepsilon_{c2}<\infty$. For the system parameter values we are working with, $\varepsilon_{c1}$ and $\varepsilon_{c2}$ are approximately $0.2$ and $4.3$ respectively.\cite{sch2015} At these two parameter values, the transverse Lyapunov exponent is zero and for $\varepsilon\in(\varepsilon_{c1},\varepsilon_{c2}),$ the exponent is negative.
Our point of interest is that the two subsystems, for $\varepsilon>\varepsilon_{c2}$, are not only not in complete synchrony but also the driven system shows unstable response and becomes unbounded. It has been analytically concluded\cite{pea1997} that if $\varepsilon$ is very large, the diffusively coupled R\"ossler subsystems can be thought of as having a driven subsystem $(y_2,y_3)$ which is being driven by the $x_1$-signal that replaces $y_1$ ({complete replacement}). This driven system has one positive conditional Lyapunov exponent making the response unstable. In effect there is no well-behaved driven attractor. Note that one is still able do a numerical calculation of the transverse Lyapunov exponents as the equations for the transverse perturbations do not involve any variables from the driven subsystem.

\begin{figure}
  \begin{center}
    \includegraphics[width=\columnwidth]{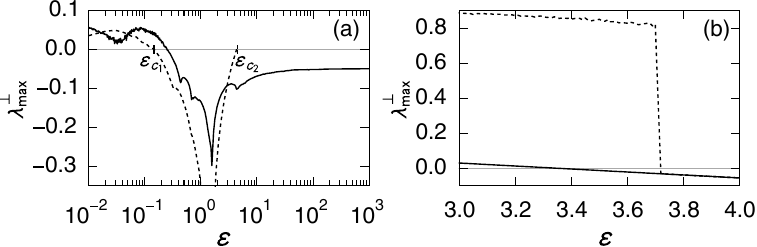}
  \end{center}
  \caption{{Transient uncoupling induces generalized synchronization.} Here we plot the maximum transverse Lyapunov exponent $\lambda^{\perp}_{\text{max}}$ as a function of the coupling strength $\varepsilon$ for (a) the R\"{o}ssler and transformed R\"{o}ssler system and (b) the dynamo--Lorenz system, without (dashed line) and with transient uncoupling (solid line). Evidently, for both the systems with transient uncoupling employed, generalized synchronization states have been induced for the coupling parameters at which they are not realizable otherwise.}
      \label{fig:unidi}
  \end{figure}

Now, if one attempts to find generalized synchronization between the R\"ossler oscillator and the transformed R\"ossler oscillator for $\varepsilon>\varepsilon_{c2}$, then one never succeeds as there is no overall chaotic attractor on which a functional relation between $\mathbf{x}$ and $\mathbf{z}$ exists. Consequently, as seen in Fig.~\ref{fig:unidi}(a), the graph of the transverse Lyapunov exponent exponent vs. coupling parameter has not been extended beyond $\varepsilon_{c2}$.

While many choices of uncoupling region are possible, let us observe what happens for a specific one: $\mathbb{U} = \{(z_1,z_2,z_3) \in \mathbb{R}^3 \mid z_1 \notin [-4.0, +6.4]\}$. A close inspection of Fig.~\ref{fig:unidi}(a) reveals that transient uncoupling increases the range of coupling parameter values for which the subsystems are in generalized synchrony with each other. Generalized synchronization is preserved even beyond $\epsilon_{c2} \approx 4.3$ for arbitrarily high values of coupling strengths. In this context, it is worth noting that the abscissa of the figure is in $\log$-scale.

In passing, it may be mentioned that the above discussion has nothing to do with the noninvertibility of transformation (\ref{eq:tn}). Even a linear (hence invertible) transformation would not have necessarily resulted in generalized synchronization because mere equivalence between the drive and the driven does not guarantee it.\cite{koc1996}
\subsection{The dynamo--Lorenz system}

The preceding example has demonstrated how the idea that transient uncoupling improves complete synchronization naturally carries over to the improvement of generalized synchronization due to transient uncoupling. Now we focus on the dynamo--Lorenz system in which we have no a priori knowledge of the functional relationship existing between the two subsystems. Thus, only numerical experiments reveal whether transient uncoupling is of any help in inducing synchrony in this system of nonidentical chaotic subsystems.

The dynamo subsystem is described by
   \begin{subequations} \label{eq:Dynamo}
    \begin{align}
      \dot{x}_1 &= x_2x_3 - \mu x_1\,, \label{eq:Dynamo_a} \\
      \dot{x}_2 &=~(x_3-\gamma)x_1 - \mu x_2\,, \label{eq:Dynamo_b} \\
      \dot{x}_3 &= 1 - x_1x_2\,; \label{eq:Dynamo_c}
    \end{align}
  \end{subequations}
  and the Lorenz subsystem by
   \begin{subequations} \label{eq:Lorenz}
    \begin{align}
      \dot{y}_1 &= \sigma(y_2-y_1)\,,\label{eq:Lorenz_a} \\
      \dot{y}_2 &= \rho y_1-y_2-y_1y_3\,, \label{eq:Lorenz_b} \\
      \dot{y}_3 &= y_1y_2-\beta y_3\,. \label{eq:Lorenz_c}
    \end{align}
  \end{subequations}
  We set $\mu = 1.7$,~$\gamma = 0.5$,~$\sigma=10$,~$\rho=35$, and~$\beta=8/3$.
    The Lorenz subsystem (driven) is coupled to the dynamo subsystem (driver) unidirectionally by the~$3 \times 3$ coupling matrix~$C$ with~$C_{ij}=1$ for~$i=j=1$ and~$C_{ij}=0$ otherwise. This corresponds to adding a term~$\varepsilon(x_1 - y_1)$ to Eq.~(\ref{eq:Lorenz_a}). At~$\varepsilon=0$ the two subsystems are not in synchrony. However, as $\varepsilon$ is increased beyond~$\varepsilon \approx 3.7$, the driven Lorenz unit enters into a generalized synchronization state with respect to the driving dynamo unit.

Among many choices of the uncoupling region $\mathbb{U}$, we find that the region defined by $y_3<33$ in the phase space of the driven unit [vide Fig.~\ref{fig:unidi}(b)] serves as a favorable uncoupling region: the transient uncoupling brings about generalized synchronization at smaller values of $\varepsilon$. Specifically, now generalized synchronization states are observed for $\varepsilon\gtrsim 3.4$.

The studies described above have showcased the possibility that in \emph{arbitrary}~systems, consisting of coupled nonidentical units in a generalized synchronization state, transient uncoupling augments the range of coupling parameter over which the generalized synchronization states are stable. In fact, further investigations show that similar conclusions hold for networks of nonidentical oscillators which we shall report elsewhere. However, an important aspect, or rather limitation, of transient uncoupling is that it is not known beforehand which uncoupling regions induces generalized synchronization. On the basis of the numerical experiments performed on the two representative systems, it may appear that choosing a $\mathbb{U}$ that assists in realizing generalized synchronization is practically a matter of trial-and-error.
\section{Why does uncoupling work?}
\begin{figure*}[t]
    \includegraphics[width=0.9\textwidth]{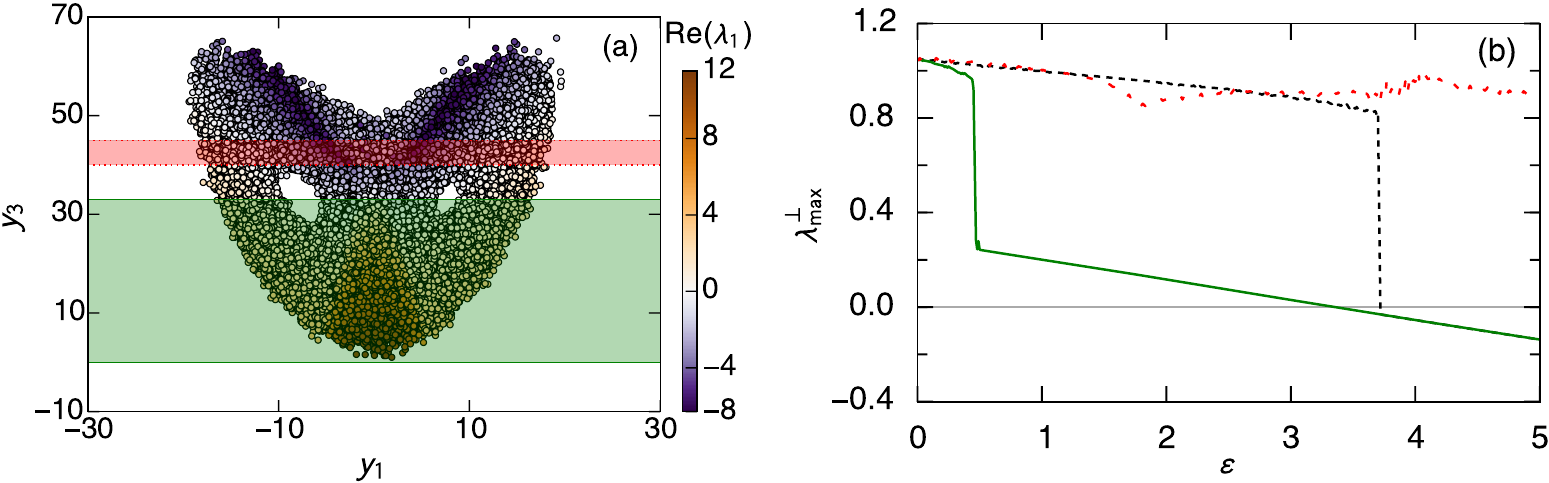}
      \caption{\emph{(Color online)} {Choosing uncoupling region to induce generalized synchronization.} Subplot (a) depicts the projected phase portrait of the Lorenz subsystem's attractor when driven by the dynamo subsystem. The colorbar encodes~$\textrm{Re}(\lambda_1)$ (vide text) of the corresponding transverse systems. The green and the red shaded areas showcase the chosen uncoupling regions---$\mathbb{U} = \{(y_1,y_2,y_3) \in \mathbb{R}^3 \mid 40<y_3<45\}$ and $\mathbb{U} = \{(y_1,y_2,y_3) \in \mathbb{R}^3 \mid y_3<33\}$ respectively---used to test the effect of transient uncoupling. Subplot (b) shows the variation of the maximal transverse Lyapunov exponents $\lambda_{\rm max}^{\bot}$ as a function of the coupling strength $\varepsilon$. The dotted red and the solid green curves respectively correspond to the red and the green uncoupling regions (as illustrated in the left panel). Their comparison with the case where transient uncoupling is not active (dashed black curve), shows how an educated choice of uncoupling region enhances generalized synchronization. Fig.~\ref{fig:unidi}b exhibits the enlarged version of the subplot near $\varepsilon=3.4$.}
    \label{fig:unidiwhy}
\end{figure*}
Note that unlike the investigation of synchronization among identical subsystems, we are now dealing with three subsystems viz.~$X$ (driven),~$Y$ (driver), and~$Y'$ (auxiliary of driven). Therefore, it is a practical question to ask which of the three subsystems should be used for defining the uncoupling region~$\mathbb{U}$ in Eq.~(\ref{eq:TUS}). More importantly, as pointed out earlier, it is not a priori clear which particular choice of~$\mathbb{U}$ is favorable as far as supporting a synchronized state is concerned.

In order to address these issues, we return to Eq.~(\ref{eq:Transverse}) and define~$\tilde{\mathbf{J}}(\mathbf{y}_s,\varepsilon) =: \mathbf{J}(\mathbf{y}_s) - \varepsilon C$. The linear stability of the synchronous state~$\mathbf{y}_s(t)$ is dictated by the largest eigenvalue of the matrix~$\tilde{\mathbf{J}}(\mathbf{y}_s,\varepsilon)$ which is a function of the variables of the driven subsystem~$\mathbf{y}_s$. Therefore, it makes sense to utilize the phase space of driven subsystem~$Y$ to pick the uncoupling region. In fact, in the aforementioned numerical experiment, this is exactly what has been done.

Now coming to the central issue, it is desirable to devise an algorithm for picking a favorable uncoupling region so that synchrony is realized for the widest possible range of coupling strengths. It would be even better if the algorithm is applicable independent of system under consideration. Although such a universal algorithm is not available to us presently, we present an intuitive strategy that explains what kind of uncoupling region may effect generalized synchronization:  Let the eigenvalues of~$\tilde{\mathbf{J}}(\mathbf{y}_s,\varepsilon)$ be given by~$\lambda_1,\lambda_2,\cdots,\lambda_N$ such that~$\textrm{Re}(\lambda_1)\ge\textrm{Re}(\lambda_2)\ge \cdots \ge\textrm{Re}(\lambda_N)$. Suppose there exist regions in the phase space of~$Y$ such that~$\textrm{Re}(\lambda_1)>0$. The points of the synchronous trajectory in these regions are linearly unstable. In order to get rid of such regions having obvious destabilizing effect, we propose to uncouple the subsystems when the phase orbits are in these regions. In place of this convenient usage of eigenvalues of~$\tilde{\mathbf{J}}(\mathbf{y}_s,\varepsilon)$, one could also have used qualitatively equivalent quantities like local Lyapunov exponents\cite{abk1991,eck1993} or eigenvalues of the symmetrized~$\tilde{\mathbf{J}}(\mathbf{y}_s,\varepsilon)$\cite{doe1991} in order to locate an optimum uncoupling region. However, our choice is sufficient for the problem in hand. In fact, Johnson and collaborators\cite{jmcp98} have also preferred the same choice to the others for exploring synchronization and imposed bifurcations in the presence of large parameter mismatch between the drive and the driven subsystems.

To explain the mechanism behind this strategy, we remark that in general the dynamics is locally contracting or locally diverging, depending on the state $(\mathbf{x}(t), \mathbf{y}(t)) \in (X,Y)$ along its trajectory determined by Eq.~(\ref{eq:FG}). By definition, the driven original dynamics $\mathbf{y}(t)$ and auxiliary dynamics $\mathbf{y}'(t)$
desynchronize when the maximum transverse Lyapunov
exponent is positive and synchronize when the exponent
is negative. But these two subsystems $Y$ and $Y'$ are identical and
uncoupled by construction, and both are forced by $X$. Consequently, for $\mathbf{y}_\bot(0) \rightarrow 0$, the
maximum transverse Lyapunov exponent is determined
in the transverse manifold of $Y\oplus Y'$, exactly as if one is
calculating normal maximum Lyapunov exponent (conditioned on $\mathbf{x}(t)$)
of the corresponding orbit $\mathbf{y}(t)$ in the phase space
of $Y$. Now, the maximum transverse Lyapunov exponent represents
the cumulative effect of all possible local contractions
and divergences along this trajectory.
So by uncoupling the two subsystems at a certain point $\mathbf{y}$, if the
local linear instability (divergence) due to the presence
of an unstable eigenspace is suppressed, then the maximum
transverse Lyapunov exponent becomes smaller. Thus,
a set of such points $\mathbf{y}$ where coupling is turned off constitutes
an uncoupling region that makes the exponent
negative, resulting in generalized synchronization.

Equipped with this method of selecting the uncoupling region, we now revisit the dynamo--Lorenz system and demonstrate why transient uncoupling is effective in enhancing generalized synchronization therein. To this end, Fig.~\ref{fig:unidiwhy} is self-explanatory: one notes that picking an uncoupling region $\mathbb{U}$ such that $\textrm{Re}(\lambda_1)>0$ for most of the phase points in the region induces synchronization. We also see that on choosing $\mathbb{U}$ such that $\textrm{Re}(\lambda_1)<0$ for a majority of its points, transient uncoupling desynchronizes the coupled system that is otherwise in generalized synchrony.

Although the aforementioned method of selecting $\mathbb{U}$ has worked remarkably well for the dynamo--Lorenz system, it is too elementary to work for every possible system. When transient uncoupling is employed, one effectively replaces the matrix~$\tilde{\mathbf{J}}(\mathbf{y}_s,\varepsilon)$ having eigenvalues~$\lambda_1, \lambda_2, \cdots, \lambda_N$ with the matrix~$\tilde{\mathbf{J}}(\mathbf{y}_s,0)$ having eigenvalues~$\lambda^0_1, \lambda^0_2, \cdots, \lambda^0_N$ (say). Therefore, if at some point of the driven phase space $\textrm{Re}(\lambda^0_1) < \textrm{Re}(\lambda_1)$, then the stability of the synchronization state improves locally. Nevertheless, even though we may, by design, choose the uncoupling region so that~$\textrm{Re}(\lambda_1)$ is a large positive number, it is entirely possible that transient uncoupling will fail to work. This is simply because it may happen that $\textrm{Re}(\lambda^0_1) > \textrm{Re}(\lambda_1)$ for a majority of points in the uncoupling region. Additionally, this methodology is not accounting for the effects of $\lambda_i$'s ($1<i<N$) as well.
\section{uncoupling overcomes noise}
Up to now we have dealt with transient uncoupling in idealized systems in the absence of any noise.
However, in realistic systems noise is always expected to be present in some form or the other.
Furthermore, low-dimensional deterministic models of natural systems do not fully describe all the external and internal fluctuations the system components are subjected to.
A common way of accounting for some of the effects of such stochastic variabilities is to add external additive noise sources into the system's evolution equations.
Interestingly, noise modeled in this fashion is known to play a crucial role in either disrupting or, more surprisingly, helping to achieve synchronized states.
An elaborate discussion regarding the debates on noise induced synchronization can be found in the review article by Boccaletti and coauthors.\cite{bocc02}

In view of the above, exploring the interplay between transient uncoupling and noise is a natural next step of our investigation. After all, both transient uncoupling and noise are observed to modify the critical value of coupling strength at which transition from synchronous to non-synchronous state occurs. For the dynamo-Lorenz system, the direction of coupling determines the role additive noise plays in either enhancing or destroying generalized synchronization:\cite{gll2006} specifically, when the Lorenz subsystem is driven by the dynamo subsystem, strong enough common noise destroys generalized synchronization. In what follows we discover that the desynchronization induced by noise might be removed by transient uncoupling.

\begin{figure}[h]

  \includegraphics[width=\columnwidth]{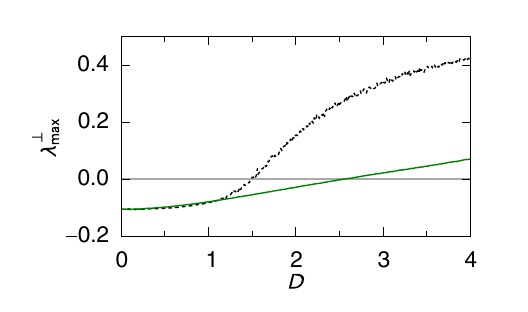}

  \caption{\emph{(Color online)} {Transient uncoupling removes noise induced desynchronization.} Here we plot the maximum transverse Lyapunov exponent $\lambda^{\perp}_{\text{max}}$ as a function of the noise strength $D$ for the dynamo--Lorenz system without (dashed black) and with transient uncoupling (solid green).  The uncoupling region used is the one highlighted with green in Fig.~{\ref{fig:unidiwhy}}(a). Generalized synchronization is seen to be maintained for larger noise strengths only when the transient uncoupling is in action.}

  \label{fig:uni}

  \end{figure}

  Let the Lorenz subsystem be coupled to the dynamo unidirectionally the way explained in Sec.~IIIB with the coupling strength set to~$\varepsilon=5.2$. At this coupling strength the system exhibits generalized synchrony in the absence of noise as shown in Fig.~\ref{fig:uni}. Now suppose a common noise term~$D \eta(t)$ is added to Eqs.~(\ref{eq:Dynamo_b}) and~(\ref{eq:Lorenz_b}). Here~$D$ is the noise amplitude and~$\eta(t)$ is a Gaussian white noise signal of unit variance. As the noise strength~$D$ is increased from zero, the maximal transverse Lyapunov exponent increases and at~$D \approx 1.4$ and beyond the two subsystems are no longer in synchrony~(Fig.~\ref{fig:uni}). However, on transiently uncoupling the subsystems, we immediately witness an increase in the threshold of maximum noise below which the generalized synchronized state is stable.

  As depicted in Fig.~\ref{fig:uni}, by choosing the green uncoupling region~$\mathbb{U}$ ($z<33$) illustrated in Fig.~\ref{fig:unidi}(a), it is possible to keep the generalized synchronization state stable up to a maximum noise amplitude of~$D \approx 2.5$. It should be noted that this noise amplitude is greater than the aforementioned threshold value of $D\approx1.4$ in the absence of transient uncoupling. This clearly highlights the definite increase in the robustness of the generalized synchronization state against the disruptive effects of the noise.

  That transient uncoupling in fact acts subtly but noticeably on the synchronized states of coupled noisy subsystems is not hard to envisage. Transient uncoupling effectively decouples the driven subsystem for a finite period of time, and consequently suppresses the effect of noise from entering into it via the variables of the driver. Also, it is well known\cite{sam2014} that noise essentially changes the system parameters to new renormalized values. These renormalized parameters are, in principle, subject to further modification on uncoupling the subsystems and the final modified parameter values may be such that for those values the subsystems are known to synchronize.
\section{Discussions and Conclusions}
  Summarizing, we have shown that by transiently switching off coupling between two unidirectionally coupled nonidentical unsynchronized subsystems, it is possible to induce generalized synchronization between them. A working methodology, based on the eigenvalues of the Jacobian of the transverse system, for selecting favorable uncoupling regions in phase space has been proven to be effective in synchronizing the dynamo--Lorenz system. Moreover, we have shown that transient uncoupling is a robust scheme for synchronization in realistic settings. It counters the noise that disrupts synchrony between two chaotic oscillators. Although with a view to avoiding repetition of qualitatively similar results we have not explicitly reported the effectiveness of transient uncoupling in synchronizing extended systems, it may be naturally and correctly guessed that transient uncoupling should also be able to synchronize bidirectionally coupled subsystems and networks of coupled chaotic nonidentical subsystems. It could also be of help in secure communication schemes based on chaotic synchronization.\cite{co1993,kp1995} In transient uncoupling induced chaotic synchronization, the drive variables (which now could also include a signal to be masked and transmitted) are required only partially for synchronizing the receiver subsystem. Thus, it might be difficult for an eavesdropper to decrypt the signal as the information transmitted per unit time is small.

  Our specific study with the R\"ossler and transformed R\"ossler system makes it clear that apart from making an existing generalized synchronization state more stable and robust, transient uncoupling can bring an overall attractor into being for the parameter values at which it is nonexistent. This is crucial because the definition of generalized synchronization requires a functional (differentiable or nondifferentiable) relationship between the drive and the driven trajectories of such an attractor. Thus, generalized synchronization is not even defined in the absence of such an attractor.

  Transient uncoupling and the methodology of choosing favorable uncoupling regions may be compared with some of the other known schemes for synchronization, e.g., finite time step method,\cite{ag93} dynamic coupling,\cite{jp01} on-off coupling,\cite{cqh09} etc. Although similar in spirit, the method of transient uncoupling is fundamentally different from such synchronization techniques and, moreover, the main results of this paper concern generalized synchronization that has not been addressed at all in the context of those schemes. Also, the robustness of such schemes against the disruptive effect of noise on generalized synchronization is equally unexplored.

  The emphasis of the article has been on how transient uncoupling may induce synchronization, but there is another obvious aspect of transient uncoupling. It might also be used to effect desynchronization. For this purpose, rather than choosing an uncoupling region that induces synchronization, one needs to choose an uncoupling region such that an already synchronized system becomes desynchronized. We have already seen such an example in Fig.~\ref{fig:unidiwhy}(b) where the uncoupling region specified by~$40<y_3<45$ (see also Fig.~\ref{fig:unidi}(a) near $\varepsilon_{c1}$) leads to desynchronization. It is well known that desynchronization,\cite{pop2011} an important phenomenon observed in, e.g.,~neuroscience and medicine, is sometimes desirable; it favorably disrupts the strong synchrony among neurons known to severely impair brain function causing, e.g.,~Parkinson's disease, epilepsy, etc.  However, whether there are any natural or biological systems where transient uncoupling is in action remains an open experimental question.

\acknowledgements

A.T.~and S.C.~thank Anindya Chatterjee and Prakhar Godara for helpful discussions. M.S.~acknowledges support from the G\"ottingen Graduate School for Neurosciences and Molecular Biosciences (DFG Grant GSC 226/2). Partially supported by the Federal Ministry of Education and Research (BMBF) Germany under grant no. 03SF0472E and by a grant from the Max Planck Society, both to M.T. S.C.~gratefully acknowledges financial support from the INSPIRE faculty fellowship (DST/INSPIRE/04/2013/000365) awarded by the Department of Science and Technology, India.


\begin{thebibliography}{55}%
\makeatletter
\providecommand \@ifxundefined [1]{%
 \@ifx{#1\undefined}
}%
\providecommand \@ifnum [1]{%
 \ifnum #1\expandafter \@firstoftwo
 \else \expandafter \@secondoftwo
 \fi
}%
\providecommand \@ifx [1]{%
 \ifx #1\expandafter \@firstoftwo
 \else \expandafter \@secondoftwo
 \fi
}%
\providecommand \natexlab [1]{#1}%
\providecommand \enquote  [1]{``#1''}%
\providecommand \bibnamefont  [1]{#1}%
\providecommand \bibfnamefont [1]{#1}%
\providecommand \citenamefont [1]{#1}%
\providecommand \href@noop [0]{\@secondoftwo}%
\providecommand \href [0]{\begingroup \@sanitize@url \@href}%
\providecommand \@href[1]{\@@startlink{#1}\@@href}%
\providecommand \@@href[1]{\endgroup#1\@@endlink}%
\providecommand \@sanitize@url [0]{\catcode `\\12\catcode `\$12\catcode
  `\&12\catcode `\#12\catcode `\^12\catcode `\_12\catcode `\%12\relax}%
\providecommand \@@startlink[1]{}%
\providecommand \@@endlink[0]{}%
\providecommand \url  [0]{\begingroup\@sanitize@url \@url }%
\providecommand \@url [1]{\endgroup\@href {#1}{\urlprefix }}%
\providecommand \urlprefix  [0]{URL }%
\providecommand \Eprint [0]{\href }%
\providecommand \doibase [0]{http://dx.doi.org/}%
\providecommand \selectlanguage [0]{\@gobble}%
\providecommand \bibinfo  [0]{\@secondoftwo}%
\providecommand \bibfield  [0]{\@secondoftwo}%
\providecommand \translation [1]{[#1]}%
\providecommand \BibitemOpen [0]{}%
\providecommand \bibitemStop [0]{}%
\providecommand \bibitemNoStop [0]{.\EOS\space}%
\providecommand \EOS [0]{\spacefactor3000\relax}%
\providecommand \BibitemShut  [1]{\csname bibitem#1\endcsname}%
\let\auto@bib@innerbib\@empty
%</preamble>
\bibitem [{\citenamefont {Schr\"oder}\ \emph {et~al.}(2015)\citenamefont
  {Schr\"oder}, \citenamefont {Mannattil}, \citenamefont {Dutta}, \citenamefont
  {Chakraborty},\ and\ \citenamefont {Timme}}]{sch2015}%
  \BibitemOpen
  \bibfield  {author} {\bibinfo {author} {\bibfnamefont {M.}~\bibnamefont
  {Schr\"oder}}, \bibinfo {author} {\bibfnamefont {M.}~\bibnamefont
  {Mannattil}}, \bibinfo {author} {\bibfnamefont {D.}~\bibnamefont {Dutta}},
  \bibinfo {author} {\bibfnamefont {S.}~\bibnamefont {Chakraborty}}, \ and\
  \bibinfo {author} {\bibfnamefont {M.}~\bibnamefont {Timme}},\ }\bibfield
  {title} {\enquote {\bibinfo {title} {Transient uncoupling induces
  synchronization},}\ }\href {\doibase 10.1103/PhysRevLett.115.054101}
  {\bibfield  {journal} {\bibinfo  {journal} {Phys. Rev. Lett.}\ }\textbf
  {\bibinfo {volume} {115}},\ \bibinfo {pages} {054101} (\bibinfo {year}
  {2015})}\BibitemShut {NoStop}%
\bibitem [{\citenamefont {Vicsek}\ \emph {et~al.}(1995)\citenamefont {Vicsek},
  \citenamefont {Czir\'ok}, \citenamefont {Ben-Jacob}, \citenamefont {Cohen},\
  and\ \citenamefont {Shochet}}]{vic1995}%
  \BibitemOpen
  \bibfield  {author} {\bibinfo {author} {\bibfnamefont {T.}~\bibnamefont
  {Vicsek}}, \bibinfo {author} {\bibfnamefont {A.}~\bibnamefont {Czir\'ok}},
  \bibinfo {author} {\bibfnamefont {E.}~\bibnamefont {Ben-Jacob}}, \bibinfo
  {author} {\bibfnamefont {I.}~\bibnamefont {Cohen}}, \ and\ \bibinfo {author}
  {\bibfnamefont {O.}~\bibnamefont {Shochet}},\ }\bibfield  {title} {\enquote
  {\bibinfo {title} {Novel type of phase transition in a system of self-driven
  particles},}\ }\href {\doibase 10.1103/PhysRevLett.75.1226} {\bibfield
  {journal} {\bibinfo  {journal} {Phys. Rev. Lett.}\ }\textbf {\bibinfo
  {volume} {75}},\ \bibinfo {pages} {1226} (\bibinfo {year}
  {1995})}\BibitemShut {NoStop}%
\bibitem [{\citenamefont {Blasius}, \citenamefont {Huppert},\ and\
  \citenamefont {Stone}(1999)}]{bla1999}%
  \BibitemOpen
  \bibfield  {author} {\bibinfo {author} {\bibfnamefont {B.}~\bibnamefont
  {Blasius}}, \bibinfo {author} {\bibfnamefont {A.}~\bibnamefont {Huppert}}, \
  and\ \bibinfo {author} {\bibfnamefont {L.}~\bibnamefont {Stone}},\ }\bibfield
   {title} {\enquote {\bibinfo {title} {Complex dynamics and phase
  synchronization in spatially extended ecological systems},}\ }\href {\doibase
  10.1038/20676} {\bibfield  {journal} {\bibinfo  {journal} {Nature}\ }\textbf
  {\bibinfo {volume} {399}},\ \bibinfo {pages} {354} (\bibinfo {year}
  {1999})}\BibitemShut {NoStop}%
\bibitem [{\citenamefont {Stone}\ \emph {et~al.}(2002)\citenamefont {Stone},
  \citenamefont {Olinky}, \citenamefont {Blasius}, \citenamefont {Huppert},\
  and\ \citenamefont {Cazelles}}]{lew2002}%
  \BibitemOpen
  \bibfield  {author} {\bibinfo {author} {\bibfnamefont {L.}~\bibnamefont
  {Stone}}, \bibinfo {author} {\bibfnamefont {R.}~\bibnamefont {Olinky}},
  \bibinfo {author} {\bibfnamefont {B.}~\bibnamefont {Blasius}}, \bibinfo
  {author} {\bibfnamefont {A.}~\bibnamefont {Huppert}}, \ and\ \bibinfo
  {author} {\bibfnamefont {B.}~\bibnamefont {Cazelles}},\ }\bibfield  {title}
  {\enquote {\bibinfo {title} {Complex synchronization phenomena in ecological
  systems},}\ }\href {\doibase 10.1063/1.1487695} {\bibfield  {journal}
  {\bibinfo  {journal} {AIP Conf. Proc.}\ }\textbf {\bibinfo {volume} {622}},\
  \bibinfo {pages} {476} (\bibinfo {year} {2002})}\BibitemShut {NoStop}%
\bibitem [{\citenamefont {Aguiar}\ \emph {et~al.}(2011)\citenamefont {Aguiar},
  \citenamefont {Ashwin}, \citenamefont {Dias},\ and\ \citenamefont
  {Field}}]{agu2011}%
  \BibitemOpen
  \bibfield  {author} {\bibinfo {author} {\bibfnamefont {M.}~\bibnamefont
  {Aguiar}}, \bibinfo {author} {\bibfnamefont {P.}~\bibnamefont {Ashwin}},
  \bibinfo {author} {\bibfnamefont {A.}~\bibnamefont {Dias}}, \ and\ \bibinfo
  {author} {\bibfnamefont {M.}~\bibnamefont {Field}},\ }\bibfield  {title}
  {\enquote {\bibinfo {title} {Dynamics of coupled cell networks: Synchrony,
  heteroclinic cycles and inflation},}\ }\href {\doibase
  10.1007/s00332-010-9083-9} {\bibfield  {journal} {\bibinfo  {journal} {J.
  Nonlin. Sci.}\ }\textbf {\bibinfo {volume} {21}},\ \bibinfo {pages} {271}
  (\bibinfo {year} {2011})}\BibitemShut {NoStop}%
\bibitem [{\citenamefont {Timme}(2006)}]{tim2006}%
  \BibitemOpen
  \bibfield  {author} {\bibinfo {author} {\bibfnamefont {M.}~\bibnamefont
  {Timme}},\ }\bibfield  {title} {\enquote {\bibinfo {title} {Does dynamics
  reflect topology in directed networks?}}\ }\href {\doibase
  10.1209/epl/i2006-10289-y} {\bibfield  {journal} {\bibinfo  {journal} {EPL}\
  }\textbf {\bibinfo {volume} {76}},\ \bibinfo {pages} {367} (\bibinfo {year}
  {2006})}\BibitemShut {NoStop}%
\bibitem [{\citenamefont {Steingrube}\ \emph {et~al.}(2010)\citenamefont
  {Steingrube}, \citenamefont {Timme}, \citenamefont {W\"{o}rg\"{o}ter},\ and\
  \citenamefont {Manoonpong}}]{stw2010}%
  \BibitemOpen
  \bibfield  {author} {\bibinfo {author} {\bibfnamefont {S.}~\bibnamefont
  {Steingrube}}, \bibinfo {author} {\bibfnamefont {M.}~\bibnamefont {Timme}},
  \bibinfo {author} {\bibfnamefont {F.}~\bibnamefont {W\"{o}rg\"{o}ter}}, \
  and\ \bibinfo {author} {\bibfnamefont {P.}~\bibnamefont {Manoonpong}},\
  }\bibfield  {title} {\enquote {\bibinfo {title} {Self-organized adaptation of
  a simple neural circuit enables complex robot behaviour},}\ }\href {\doibase
  10.1038/nphys1508} {\bibfield  {journal} {\bibinfo  {journal} {Nat. Phys.}\
  }\textbf {\bibinfo {volume} {6}},\ \bibinfo {pages} {224} (\bibinfo {year}
  {2010})}\BibitemShut {NoStop}%
\bibitem [{\citenamefont {Tanaka}(2014)}]{tan2014}%
  \BibitemOpen
  \bibfield  {author} {\bibinfo {author} {\bibfnamefont {H.-A.}\ \bibnamefont
  {Tanaka}},\ }\bibfield  {title} {\enquote {\bibinfo {title} {Synchronization
  limit of weakly forced nonlinear oscillators},}\ }\href {\doibase
  10.1088/1751-8113/47/40/402002} {\bibfield  {journal} {\bibinfo  {journal}
  {J. Phys. A}\ }\textbf {\bibinfo {volume} {47}},\ \bibinfo {pages} {402002}
  (\bibinfo {year} {2014})}\BibitemShut {NoStop}%
\bibitem [{\citenamefont {Matheny}\ \emph {et~al.}(2014)\citenamefont
  {Matheny}, \citenamefont {Grau}, \citenamefont {Villanueva}, \citenamefont
  {Karabalin}, \citenamefont {Cross},\ and\ \citenamefont {Roukes}}]{mea2014}%
  \BibitemOpen
  \bibfield  {author} {\bibinfo {author} {\bibfnamefont {M.~H.}\ \bibnamefont
  {Matheny}}, \bibinfo {author} {\bibfnamefont {M.}~\bibnamefont {Grau}},
  \bibinfo {author} {\bibfnamefont {L.~G.}\ \bibnamefont {Villanueva}},
  \bibinfo {author} {\bibfnamefont {R.~B.}\ \bibnamefont {Karabalin}}, \bibinfo
  {author} {\bibfnamefont {M.~C.}\ \bibnamefont {Cross}}, \ and\ \bibinfo
  {author} {\bibfnamefont {M.~L.}\ \bibnamefont {Roukes}},\ }\bibfield  {title}
  {\enquote {\bibinfo {title} {Phase synchronization of two anharmonic
  nanomechanical oscillators},}\ }\href {\doibase
  10.1103/PhysRevLett.112.014101} {\bibfield  {journal} {\bibinfo  {journal}
  {Phys. Rev. Lett.}\ }\textbf {\bibinfo {volume} {112}},\ \bibinfo {pages}
  {014101} (\bibinfo {year} {2014})}\BibitemShut {NoStop}%
\bibitem [{\citenamefont {Strogatz}\ \emph {et~al.}(2005)\citenamefont
  {Strogatz}, \citenamefont {Abrams}, \citenamefont {McRobie}, \citenamefont
  {Eckhardt},\ and\ \citenamefont {Ott}}]{str2005}%
  \BibitemOpen
  \bibfield  {author} {\bibinfo {author} {\bibfnamefont {S.~H.}\ \bibnamefont
  {Strogatz}}, \bibinfo {author} {\bibfnamefont {D.~M.}\ \bibnamefont
  {Abrams}}, \bibinfo {author} {\bibfnamefont {A.}~\bibnamefont {McRobie}},
  \bibinfo {author} {\bibfnamefont {B.}~\bibnamefont {Eckhardt}}, \ and\
  \bibinfo {author} {\bibfnamefont {E.}~\bibnamefont {Ott}},\ }\bibfield
  {title} {\enquote {\bibinfo {title} {Theoretical mechanics: Crowd synchrony
  on the {Millennium Bridge}},}\ }\href {\doibase 10.1038/43843a} {\bibfield
  {journal} {\bibinfo  {journal} {Nature}\ }\textbf {\bibinfo {volume} {438}},\
  \bibinfo {pages} {43} (\bibinfo {year} {2005})}\BibitemShut {NoStop}%
\bibitem [{\citenamefont {Pikovsky}, \citenamefont {Rosenblum},\ and\
  \citenamefont {Kurths}(2001)}]{prk2003}%
  \BibitemOpen
  \bibfield  {author} {\bibinfo {author} {\bibfnamefont {A.}~\bibnamefont
  {Pikovsky}}, \bibinfo {author} {\bibfnamefont {M.}~\bibnamefont {Rosenblum}},
  \ and\ \bibinfo {author} {\bibfnamefont {J.}~\bibnamefont {Kurths}},\
  }\href@noop {} {\emph {\bibinfo {title} {Synchronization}}}\ (\bibinfo
  {publisher} {Cambridge University Press, New York},\ \bibinfo {year}
  {2001})\BibitemShut {NoStop}%
\bibitem [{\citenamefont {Balanov}\ \emph {et~al.}(2010)\citenamefont
  {Balanov}, \citenamefont {Janson}, \citenamefont {Postnov},\ and\
  \citenamefont {Sosnovtseva}}]{bjps2009}%
  \BibitemOpen
  \bibfield  {author} {\bibinfo {author} {\bibfnamefont {A.}~\bibnamefont
  {Balanov}}, \bibinfo {author} {\bibfnamefont {N.}~\bibnamefont {Janson}},
  \bibinfo {author} {\bibfnamefont {D.}~\bibnamefont {Postnov}}, \ and\
  \bibinfo {author} {\bibfnamefont {O.}~\bibnamefont {Sosnovtseva}},\
  }\href@noop {} {\emph {\bibinfo {title} {Synchronization}}}\ (\bibinfo
  {publisher} {Springer, Berlin},\ \bibinfo {year} {2010})\BibitemShut
  {NoStop}%
\bibitem [{\citenamefont {Yamada}\ and\ \citenamefont
  {Fujisaka}(1983)}]{yam1983}%
  \BibitemOpen
  \bibfield  {author} {\bibinfo {author} {\bibfnamefont {T.}~\bibnamefont
  {Yamada}}\ and\ \bibinfo {author} {\bibfnamefont {H.}~\bibnamefont
  {Fujisaka}},\ }\bibfield  {title} {\enquote {\bibinfo {title} {Stability
  theory of synchronized motion in coupled-oscillator systems. ii: The mapping
  approach},}\ }\href {\doibase 10.1143/PTP.70.1240} {\bibfield  {journal}
  {\bibinfo  {journal} {Prog. Theor. Phys.}\ }\textbf {\bibinfo {volume}
  {70}},\ \bibinfo {pages} {1240} (\bibinfo {year} {1983})}\BibitemShut
  {NoStop}%
\bibitem [{\citenamefont {Pecora}\ and\ \citenamefont
  {Carroll}(1990)}]{pc1990}%
  \BibitemOpen
  \bibfield  {author} {\bibinfo {author} {\bibfnamefont {L.~M.}\ \bibnamefont
  {Pecora}}\ and\ \bibinfo {author} {\bibfnamefont {T.~L.}\ \bibnamefont
  {Carroll}},\ }\bibfield  {title} {\enquote {\bibinfo {title} {Synchronization
  in chaotic systems},}\ }\href {\doibase 10.1103/PhysRevLett.64.821}
  {\bibfield  {journal} {\bibinfo  {journal} {Phys. Rev. Lett.}\ }\textbf
  {\bibinfo {volume} {64}},\ \bibinfo {pages} {821} (\bibinfo {year}
  {1990})}\BibitemShut {NoStop}%
\bibitem [{\citenamefont {Rosenblum}, \citenamefont {Pikovsky},\ and\
  \citenamefont {Kurths}(1996)}]{rpk1996}%
  \BibitemOpen
  \bibfield  {author} {\bibinfo {author} {\bibfnamefont {M.~G.}\ \bibnamefont
  {Rosenblum}}, \bibinfo {author} {\bibfnamefont {A.~S.}\ \bibnamefont
  {Pikovsky}}, \ and\ \bibinfo {author} {\bibfnamefont {J.}~\bibnamefont
  {Kurths}},\ }\bibfield  {title} {\enquote {\bibinfo {title} {Phase
  synchronization of chaotic oscillators},}\ }\href {\doibase
  10.1103/PhysRevLett.76.1804} {\bibfield  {journal} {\bibinfo  {journal}
  {Phys. Rev. Lett.}\ }\textbf {\bibinfo {volume} {76}},\ \bibinfo {pages}
  {1804} (\bibinfo {year} {1996})}\BibitemShut {NoStop}%
\bibitem [{\citenamefont {Rosa}, \citenamefont {Ott},\ and\ \citenamefont
  {Hess}(1998)}]{roh1998}%
  \BibitemOpen
  \bibfield  {author} {\bibinfo {author} {\bibfnamefont {E.}~\bibnamefont
  {Rosa}}, \bibinfo {author} {\bibfnamefont {E.}~\bibnamefont {Ott}}, \ and\
  \bibinfo {author} {\bibfnamefont {M.~H.}\ \bibnamefont {Hess}},\ }\bibfield
  {title} {\enquote {\bibinfo {title} {Transition to phase synchronization of
  chaos},}\ }\href {\doibase 10.1103/PhysRevLett.80.1642} {\bibfield  {journal}
  {\bibinfo  {journal} {Phys. Rev. Lett.}\ }\textbf {\bibinfo {volume} {80}},\
  \bibinfo {pages} {1642} (\bibinfo {year} {1998})}\BibitemShut {NoStop}%
\bibitem [{\citenamefont {Izhikevich}(2000)}]{izh2000}%
  \BibitemOpen
  \bibfield  {author} {\bibinfo {author} {\bibfnamefont {E.~M.}\ \bibnamefont
  {Izhikevich}},\ }\bibfield  {title} {\enquote {\bibinfo {title} {Neural
  excitability, spiking and bursting},}\ }\href {\doibase
  10.1142/S0218127400000840} {\bibfield  {journal} {\bibinfo  {journal} {Int.
  J. Bifur. Chaos}\ }\textbf {\bibinfo {volume} {10}},\ \bibinfo {pages} {1171}
  (\bibinfo {year} {2000})}\BibitemShut {NoStop}%
\bibitem [{\citenamefont {Dhamala}, \citenamefont {Jirsa},\ and\ \citenamefont
  {Ding}(2004)}]{djd2004}%
  \BibitemOpen
  \bibfield  {author} {\bibinfo {author} {\bibfnamefont {M.}~\bibnamefont
  {Dhamala}}, \bibinfo {author} {\bibfnamefont {V.~K.}\ \bibnamefont {Jirsa}},
  \ and\ \bibinfo {author} {\bibfnamefont {M.}~\bibnamefont {Ding}},\
  }\bibfield  {title} {\enquote {\bibinfo {title} {Transitions to synchrony in
  coupled bursting neurons},}\ }\href {\doibase 10.1103/PhysRevLett.92.028101}
  {\bibfield  {journal} {\bibinfo  {journal} {Phys. Rev. Lett.}\ }\textbf
  {\bibinfo {volume} {92}},\ \bibinfo {pages} {028101} (\bibinfo {year}
  {2004})}\BibitemShut {NoStop}%
\bibitem [{\citenamefont {Rosenblum}, \citenamefont {Pikovsky},\ and\
  \citenamefont {Kurths}(1997)}]{rpk1997}%
  \BibitemOpen
  \bibfield  {author} {\bibinfo {author} {\bibfnamefont {M.~G.}\ \bibnamefont
  {Rosenblum}}, \bibinfo {author} {\bibfnamefont {A.~S.}\ \bibnamefont
  {Pikovsky}}, \ and\ \bibinfo {author} {\bibfnamefont {J.}~\bibnamefont
  {Kurths}},\ }\bibfield  {title} {\enquote {\bibinfo {title} {From phase to
  lag synchronization in coupled chaotic oscillators},}\ }\href {\doibase
  10.1103/PhysRevLett.78.4193} {\bibfield  {journal} {\bibinfo  {journal}
  {Phys. Rev. Lett.}\ }\textbf {\bibinfo {volume} {78}},\ \bibinfo {pages}
  {4193} (\bibinfo {year} {1997})}\BibitemShut {NoStop}%
\bibitem [{\citenamefont {Boccaletti}\ and\ \citenamefont
  {Valladares}(2000)}]{boc2000}%
  \BibitemOpen
  \bibfield  {author} {\bibinfo {author} {\bibfnamefont {S.}~\bibnamefont
  {Boccaletti}}\ and\ \bibinfo {author} {\bibfnamefont {D.~L.}\ \bibnamefont
  {Valladares}},\ }\bibfield  {title} {\enquote {\bibinfo {title}
  {Characterization of intermittent lag synchronization},}\ }\href {\doibase
  10.1103/PhysRevE.62.7497} {\bibfield  {journal} {\bibinfo  {journal} {Phys.
  Rev. E}\ }\textbf {\bibinfo {volume} {62}},\ \bibinfo {pages} {7497}
  (\bibinfo {year} {2000})}\BibitemShut {NoStop}%
\bibitem [{\citenamefont {Rulkov}\ \emph {et~al.}(1995)\citenamefont {Rulkov},
  \citenamefont {Sushchik}, \citenamefont {Tsimring},\ and\ \citenamefont
  {Abarbanel}}]{rul1995}%
  \BibitemOpen
  \bibfield  {author} {\bibinfo {author} {\bibfnamefont {N.~F.}\ \bibnamefont
  {Rulkov}}, \bibinfo {author} {\bibfnamefont {M.~M.}\ \bibnamefont
  {Sushchik}}, \bibinfo {author} {\bibfnamefont {L.~S.}\ \bibnamefont
  {Tsimring}}, \ and\ \bibinfo {author} {\bibfnamefont {H.~D.~I.}\ \bibnamefont
  {Abarbanel}},\ }\bibfield  {title} {\enquote {\bibinfo {title} {Generalized
  synchronization of chaos in directionally coupled chaotic systems},}\ }\href
  {\doibase 10.1103/PhysRevE.51.980} {\bibfield  {journal} {\bibinfo  {journal}
  {Phys. Rev. E}\ }\textbf {\bibinfo {volume} {51}},\ \bibinfo {pages} {980}
  (\bibinfo {year} {1995})}\BibitemShut {NoStop}%
\bibitem [{\citenamefont {Kocarev}\ and\ \citenamefont
  {Parlitz}(1996)}]{koc1996}%
  \BibitemOpen
  \bibfield  {author} {\bibinfo {author} {\bibfnamefont {L.}~\bibnamefont
  {Kocarev}}\ and\ \bibinfo {author} {\bibfnamefont {U.}~\bibnamefont
  {Parlitz}},\ }\bibfield  {title} {\enquote {\bibinfo {title} {Generalized
  synchronization, predictability, and equivalence of unidirectionally coupled
  dynamical systems},}\ }\href {\doibase 10.1103/PhysRevLett.76.1816}
  {\bibfield  {journal} {\bibinfo  {journal} {Phys. Rev. Lett.}\ }\textbf
  {\bibinfo {volume} {76}},\ \bibinfo {pages} {1816} (\bibinfo {year}
  {1996})}\BibitemShut {NoStop}%
\bibitem [{\citenamefont {Hunt}, \citenamefont {Ott},\ and\ \citenamefont
  {Yorke}(1997)}]{hoy1997}%
  \BibitemOpen
  \bibfield  {author} {\bibinfo {author} {\bibfnamefont {B.~R.}\ \bibnamefont
  {Hunt}}, \bibinfo {author} {\bibfnamefont {E.}~\bibnamefont {Ott}}, \ and\
  \bibinfo {author} {\bibfnamefont {J.~A.}\ \bibnamefont {Yorke}},\ }\bibfield
  {title} {\enquote {\bibinfo {title} {Differentiable generalized
  synchronization of chaos},}\ }\href {\doibase 10.1103/PhysRevE.55.4029}
  {\bibfield  {journal} {\bibinfo  {journal} {Phys. Rev. E}\ }\textbf {\bibinfo
  {volume} {55}},\ \bibinfo {pages} {4029} (\bibinfo {year}
  {1997})}\BibitemShut {NoStop}%
\bibitem [{\citenamefont {Pyragas}(1996)}]{pyr1996}%
  \BibitemOpen
  \bibfield  {author} {\bibinfo {author} {\bibfnamefont {K.}~\bibnamefont
  {Pyragas}},\ }\bibfield  {title} {\enquote {\bibinfo {title} {Weak and strong
  synchronization of chaos},}\ }\href {\doibase 10.1103/PhysRevE.54.R4508}
  {\bibfield  {journal} {\bibinfo  {journal} {Phys. Rev. E}\ }\textbf {\bibinfo
  {volume} {54}},\ \bibinfo {pages} {R4508} (\bibinfo {year}
  {1996})}\BibitemShut {NoStop}%
\bibitem [{\citenamefont {Brown}\ and\ \citenamefont {Kocarev}(2000)}]{bk2000}%
  \BibitemOpen
  \bibfield  {author} {\bibinfo {author} {\bibfnamefont {R.}~\bibnamefont
  {Brown}}\ and\ \bibinfo {author} {\bibfnamefont {L.}~\bibnamefont
  {Kocarev}},\ }\bibfield  {title} {\enquote {\bibinfo {title} {A unifying
  definition of synchronization for dynamical systems},}\ }\href {\doibase
  10.1063/1.166500} {\bibfield  {journal} {\bibinfo  {journal} {Chaos}\
  }\textbf {\bibinfo {volume} {10}},\ \bibinfo {pages} {344} (\bibinfo {year}
  {2000})}\BibitemShut {NoStop}%
\bibitem [{\citenamefont {Pecora}\ \emph {et~al.}(1997)\citenamefont {Pecora},
  \citenamefont {Carroll}, \citenamefont {Johnson}, \citenamefont {Mar},\ and\
  \citenamefont {Heagy}}]{pea1997}%
  \BibitemOpen
  \bibfield  {author} {\bibinfo {author} {\bibfnamefont {L.~M.}\ \bibnamefont
  {Pecora}}, \bibinfo {author} {\bibfnamefont {T.~L.}\ \bibnamefont {Carroll}},
  \bibinfo {author} {\bibfnamefont {G.~A.}\ \bibnamefont {Johnson}}, \bibinfo
  {author} {\bibfnamefont {D.~J.}\ \bibnamefont {Mar}}, \ and\ \bibinfo
  {author} {\bibfnamefont {J.~F.}\ \bibnamefont {Heagy}},\ }\bibfield  {title}
  {\enquote {\bibinfo {title} {Fundamentals of synchronization in chaotic
  systems, concepts, and applications},}\ }\href {\doibase 10.1063/1.166278}
  {\bibfield  {journal} {\bibinfo  {journal} {Chaos}\ }\textbf {\bibinfo
  {volume} {7}},\ \bibinfo {pages} {520} (\bibinfo {year} {1997})}\BibitemShut
  {NoStop}%
\bibitem [{\citenamefont {Herzel}\ and\ \citenamefont
  {Freund}(1995)}]{her1995}%
  \BibitemOpen
  \bibfield  {author} {\bibinfo {author} {\bibfnamefont {H.}~\bibnamefont
  {Herzel}}\ and\ \bibinfo {author} {\bibfnamefont {J.}~\bibnamefont
  {Freund}},\ }\bibfield  {title} {\enquote {\bibinfo {title} {Chaos, noise,
  and synchronization reconsidered},}\ }\href {\doibase
  10.1103/PhysRevE.52.3238} {\bibfield  {journal} {\bibinfo  {journal} {Phys.
  Rev. E}\ }\textbf {\bibinfo {volume} {52}},\ \bibinfo {pages} {3238}
  (\bibinfo {year} {1995})}\BibitemShut {NoStop}%
\bibitem [{\citenamefont {Lai}\ and\ \citenamefont {Zhou}(1998)}]{lai1998}%
  \BibitemOpen
  \bibfield  {author} {\bibinfo {author} {\bibfnamefont {C.-H.}\ \bibnamefont
  {Lai}}\ and\ \bibinfo {author} {\bibfnamefont {C.}~\bibnamefont {Zhou}},\
  }\bibfield  {title} {\enquote {\bibinfo {title} {Synchronization of chaotic
  maps by symmetric common noise},}\ }\href {\doibase
  10.1209/epl/i1998-00368-1} {\bibfield  {journal} {\bibinfo  {journal}
  {Europhys. Lett.}\ }\textbf {\bibinfo {volume} {43}},\ \bibinfo {pages} {376}
  (\bibinfo {year} {1998})}\BibitemShut {NoStop}%
\bibitem [{\citenamefont {Toral}\ \emph {et~al.}(2001)\citenamefont {Toral},
  \citenamefont {Mirasso}, \citenamefont {Hern{\'{a}}ndez-Garcia},\ and\
  \citenamefont {Piro}}]{rea2001}%
  \BibitemOpen
  \bibfield  {author} {\bibinfo {author} {\bibfnamefont {R.}~\bibnamefont
  {Toral}}, \bibinfo {author} {\bibfnamefont {C.~R.}\ \bibnamefont {Mirasso}},
  \bibinfo {author} {\bibfnamefont {E.}~\bibnamefont {Hern{\'{a}}ndez-Garcia}},
  \ and\ \bibinfo {author} {\bibfnamefont {O.}~\bibnamefont {Piro}},\
  }\bibfield  {title} {\enquote {\bibinfo {title} {Analytical and numerical
  studies of noise-induced synchronization of chaotic systems},}\ }\href
  {\doibase 10.1063/1.1386397} {\bibfield  {journal} {\bibinfo  {journal}
  {Chaos}\ }\textbf {\bibinfo {volume} {11}},\ \bibinfo {pages} {665} (\bibinfo
  {year} {2001})}\BibitemShut {NoStop}%
\bibitem [{\citenamefont {Wang}, \citenamefont {Chik},\ and\ \citenamefont
  {Wang}(2000)}]{wcw2000}%
  \BibitemOpen
  \bibfield  {author} {\bibinfo {author} {\bibfnamefont {Y.}~\bibnamefont
  {Wang}}, \bibinfo {author} {\bibfnamefont {D.~T.~W.}\ \bibnamefont {Chik}}, \
  and\ \bibinfo {author} {\bibfnamefont {Z.~D.}\ \bibnamefont {Wang}},\
  }\bibfield  {title} {\enquote {\bibinfo {title} {Coherence resonance and
  noise-induced synchronization in globally coupled {Hodgkin-Huxley}
  neurons},}\ }\href {\doibase 10.1103/PhysRevE.61.740} {\bibfield  {journal}
  {\bibinfo  {journal} {Phys. Rev. E}\ }\textbf {\bibinfo {volume} {61}},\
  \bibinfo {pages} {740} (\bibinfo {year} {2000})}\BibitemShut {NoStop}%
\bibitem [{\citenamefont {Zhou}\ \emph {et~al.}(2002)\citenamefont {Zhou},
  \citenamefont {Kurths}, \citenamefont {Kiss},\ and\ \citenamefont
  {Hudson}}]{zea2002}%
  \BibitemOpen
  \bibfield  {author} {\bibinfo {author} {\bibfnamefont {C.}~\bibnamefont
  {Zhou}}, \bibinfo {author} {\bibfnamefont {J.}~\bibnamefont {Kurths}},
  \bibinfo {author} {\bibfnamefont {I.~Z.}\ \bibnamefont {Kiss}}, \ and\
  \bibinfo {author} {\bibfnamefont {J.~L.}\ \bibnamefont {Hudson}},\ }\bibfield
   {title} {\enquote {\bibinfo {title} {Noise-enhanced phase synchronization of
  chaotic oscillators},}\ }\href {\doibase 10.1103/PhysRevLett.89.014101}
  {\bibfield  {journal} {\bibinfo  {journal} {Phys. Rev. Lett.}\ }\textbf
  {\bibinfo {volume} {89}},\ \bibinfo {pages} {014101} (\bibinfo {year}
  {2002})}\BibitemShut {NoStop}%
\bibitem [{\citenamefont {Zhou}\ and\ \citenamefont {Kurths}(2002)}]{zho2002}%
  \BibitemOpen
  \bibfield  {author} {\bibinfo {author} {\bibfnamefont {C.}~\bibnamefont
  {Zhou}}\ and\ \bibinfo {author} {\bibfnamefont {J.}~\bibnamefont {Kurths}},\
  }\bibfield  {title} {\enquote {\bibinfo {title} {Noise-induced phase
  synchronization and synchronization transitions in chaotic oscillators},}\
  }\href {\doibase 10.1103/PhysRevLett.88.230602} {\bibfield  {journal}
  {\bibinfo  {journal} {Phys. Rev. Lett.}\ }\textbf {\bibinfo {volume} {88}},\
  \bibinfo {pages} {230602} (\bibinfo {year} {2002})}\BibitemShut {NoStop}%
\bibitem [{\citenamefont {De~Monte}\ \emph {et~al.}(2004)\citenamefont
  {De~Monte}, \citenamefont {d'Ovidio}, \citenamefont {Chat\'e},\ and\
  \citenamefont {Mosekilde}}]{dfh2004}%
  \BibitemOpen
  \bibfield  {author} {\bibinfo {author} {\bibfnamefont {S.}~\bibnamefont
  {De~Monte}}, \bibinfo {author} {\bibfnamefont {F.}~\bibnamefont {d'Ovidio}},
  \bibinfo {author} {\bibfnamefont {H.}~\bibnamefont {Chat\'e}}, \ and\
  \bibinfo {author} {\bibfnamefont {E.}~\bibnamefont {Mosekilde}},\ }\bibfield
  {title} {\enquote {\bibinfo {title} {Noise-induced macroscopic bifurcations
  in globally coupled chaotic units},}\ }\href {\doibase
  10.1103/PhysRevLett.92.254101} {\bibfield  {journal} {\bibinfo  {journal}
  {Phys. Rev. Lett.}\ }\textbf {\bibinfo {volume} {92}},\ \bibinfo {pages}
  {254101} (\bibinfo {year} {2004})}\BibitemShut {NoStop}%
\bibitem [{\citenamefont {Koronovskii}\ \emph {et~al.}(2006)\citenamefont
  {Koronovskii}, \citenamefont {Moskalenko}, \citenamefont {Trubetskov},\ and\
  \citenamefont {Khramov}}]{kea2006}%
  \BibitemOpen
  \bibfield  {author} {\bibinfo {author} {\bibfnamefont {A.~A.}\ \bibnamefont
  {Koronovskii}}, \bibinfo {author} {\bibfnamefont {O.~I.}\ \bibnamefont
  {Moskalenko}}, \bibinfo {author} {\bibfnamefont {D.~I.}\ \bibnamefont
  {Trubetskov}}, \ and\ \bibinfo {author} {\bibfnamefont {A.~E.}\ \bibnamefont
  {Khramov}},\ }\bibfield  {title} {\enquote {\bibinfo {title} {Generalized
  synchronization and noise-induced synchronization: The same type of behavior
  of coupled chaotic systems},}\ }\href {\doibase 10.1134/S1028335806040070}
  {\bibfield  {journal} {\bibinfo  {journal} {Dokl. Phys.}\ }\textbf {\bibinfo
  {volume} {51}},\ \bibinfo {pages} {189} (\bibinfo {year} {2006})}\BibitemShut
  {NoStop}%
\bibitem [{\citenamefont {Hramov}, \citenamefont {Koronovskii},\ and\
  \citenamefont {Moskalenko}(2006)}]{hkm2006}%
  \BibitemOpen
  \bibfield  {author} {\bibinfo {author} {\bibfnamefont {A.~E.}\ \bibnamefont
  {Hramov}}, \bibinfo {author} {\bibfnamefont {A.~A.}\ \bibnamefont
  {Koronovskii}}, \ and\ \bibinfo {author} {\bibfnamefont {O.~I.}\ \bibnamefont
  {Moskalenko}},\ }\bibfield  {title} {\enquote {\bibinfo {title} {Are
  generalized synchronization and noise-induced synchronization identical types
  of synchronous behavior of chaotic oscillators?}}\ }\href {\doibase
  10.1016/j.physleta.2006.01.079} {\bibfield  {journal} {\bibinfo  {journal}
  {Phys. Lett. A}\ }\textbf {\bibinfo {volume} {354}},\ \bibinfo {pages} {423}
  (\bibinfo {year} {2006})}\BibitemShut {NoStop}%
\bibitem [{\citenamefont {Guan}, \citenamefont {Lai},\ and\ \citenamefont
  {Lai}(2006)}]{gll2006}%
  \BibitemOpen
  \bibfield  {author} {\bibinfo {author} {\bibfnamefont {S.}~\bibnamefont
  {Guan}}, \bibinfo {author} {\bibfnamefont {Y.-C.}\ \bibnamefont {Lai}}, \
  and\ \bibinfo {author} {\bibfnamefont {C.-H.}\ \bibnamefont {Lai}},\
  }\bibfield  {title} {\enquote {\bibinfo {title} {Effect of noise on
  generalized chaotic synchronization},}\ }\href {\doibase
  10.1103/PhysRevE.73.046210} {\bibfield  {journal} {\bibinfo  {journal} {Phys.
  Rev. E}\ }\textbf {\bibinfo {volume} {73}},\ \bibinfo {pages} {046210}
  (\bibinfo {year} {2006})}\BibitemShut {NoStop}%
\bibitem [{\citenamefont {Mainen}\ and\ \citenamefont
  {Sejnowski}(1995)}]{mai1995}%
  \BibitemOpen
  \bibfield  {author} {\bibinfo {author} {\bibfnamefont {Z.~F.}\ \bibnamefont
  {Mainen}}\ and\ \bibinfo {author} {\bibfnamefont {T.~J.}\ \bibnamefont
  {Sejnowski}},\ }\bibfield  {title} {\enquote {\bibinfo {title} {Reliability
  of spike timing in neocortical neurons},}\ }\href {\doibase
  10.1126/science.7770778} {\bibfield  {journal} {\bibinfo  {journal}
  {Science}\ }\textbf {\bibinfo {volume} {268}},\ \bibinfo {pages} {1503}
  (\bibinfo {year} {1995})}\BibitemShut {NoStop}%
\bibitem [{\citenamefont {He}, \citenamefont {Stone},\ and\ \citenamefont
  {Cazelles}(2010)}]{hsc2010}%
  \BibitemOpen
  \bibfield  {author} {\bibinfo {author} {\bibfnamefont {D.}~\bibnamefont
  {He}}, \bibinfo {author} {\bibfnamefont {L.}~\bibnamefont {Stone}}, \ and\
  \bibinfo {author} {\bibfnamefont {B.}~\bibnamefont {Cazelles}},\ }\bibfield
  {title} {\enquote {\bibinfo {title} {Noise-induced synchronization in
  multitrophic chaotic ecological systems},}\ }\href {\doibase
  10.1142/S0218127410026824} {\bibfield  {journal} {\bibinfo  {journal} {Int.
  J. Bifur. Chaos}\ }\textbf {\bibinfo {volume} {20}},\ \bibinfo {pages} {1779}
  (\bibinfo {year} {2010})}\BibitemShut {NoStop}%
\bibitem [{\citenamefont {Moskalenko}\ \emph {et~al.}(2012)\citenamefont
  {Moskalenko}, \citenamefont {Koronovskii}, \citenamefont {Hramov},\ and\
  \citenamefont {Boccaletti}}]{mos2012}%
  \BibitemOpen
  \bibfield  {author} {\bibinfo {author} {\bibfnamefont {O.~I.}\ \bibnamefont
  {Moskalenko}}, \bibinfo {author} {\bibfnamefont {A.~A.}\ \bibnamefont
  {Koronovskii}}, \bibinfo {author} {\bibfnamefont {A.~E.}\ \bibnamefont
  {Hramov}}, \ and\ \bibinfo {author} {\bibfnamefont {S.}~\bibnamefont
  {Boccaletti}},\ }\bibfield  {title} {\enquote {\bibinfo {title} {Generalized
  synchronization in mutually coupled oscillators and complex networks},}\
  }\href {\doibase 10.1103/PhysRevE.86.036216} {\bibfield  {journal} {\bibinfo
  {journal} {Phys. Rev. E}\ }\textbf {\bibinfo {volume} {86}},\ \bibinfo
  {pages} {036216} (\bibinfo {year} {2012})}\BibitemShut {NoStop}%
\bibitem [{\citenamefont {Abarbanel}, \citenamefont {Rulkov},\ and\
  \citenamefont {Sushchik}(1996)}]{aba1996}%
  \BibitemOpen
  \bibfield  {author} {\bibinfo {author} {\bibfnamefont {H.~D.~I.}\
  \bibnamefont {Abarbanel}}, \bibinfo {author} {\bibfnamefont {N.~F.}\
  \bibnamefont {Rulkov}}, \ and\ \bibinfo {author} {\bibfnamefont {M.~M.}\
  \bibnamefont {Sushchik}},\ }\bibfield  {title} {\enquote {\bibinfo {title}
  {Generalized synchronization of chaos: The auxiliary system approach},}\
  }\href {\doibase 10.1103/PhysRevE.53.4528} {\bibfield  {journal} {\bibinfo
  {journal} {Phys. Rev. E}\ }\textbf {\bibinfo {volume} {53}},\ \bibinfo
  {pages} {4528} (\bibinfo {year} {1996})}\BibitemShut {NoStop}%
\bibitem [{\citenamefont {Lorenz}(1963)}]{lor1963}%
  \BibitemOpen
  \bibfield  {author} {\bibinfo {author} {\bibfnamefont {E.~N.}\ \bibnamefont
  {Lorenz}},\ }\bibfield  {title} {\enquote {\bibinfo {title} {Deterministic
  nonperiodic flow},}\ }\href {\doibase 10.1007/978-0-387-21830-4_2} {\bibfield
   {journal} {\bibinfo  {journal} {J. Atmos. Sci.}\ }\textbf {\bibinfo {volume}
  {20}},\ \bibinfo {pages} {130} (\bibinfo {year} {1963})}\BibitemShut
  {NoStop}%
\bibitem [{\citenamefont {Mainieri}\ and\ \citenamefont
  {Rehacek}(1999)}]{mai1999}%
  \BibitemOpen
  \bibfield  {author} {\bibinfo {author} {\bibfnamefont {R.}~\bibnamefont
  {Mainieri}}\ and\ \bibinfo {author} {\bibfnamefont {J.}~\bibnamefont
  {Rehacek}},\ }\bibfield  {title} {\enquote {\bibinfo {title} {Projective
  synchronization in three-dimensional chaotic systems},}\ }\href {\doibase
  10.1103/PhysRevLett.82.3042} {\bibfield  {journal} {\bibinfo  {journal}
  {Phys. Rev. Lett.}\ }\textbf {\bibinfo {volume} {82}},\ \bibinfo {pages}
  {3042} (\bibinfo {year} {1999})}\BibitemShut {NoStop}%
\bibitem [{\citenamefont {R\"ossler}(1976)}]{roessler76}%
  \BibitemOpen
  \bibfield  {author} {\bibinfo {author} {\bibfnamefont {O.~E.}\ \bibnamefont
  {R\"ossler}},\ }\bibfield  {title} {\enquote {\bibinfo {title} {An equation
  for continuous chaos},}\ }\href {\doibase 10.1016/0375-9601(76)90101-8}
  {\bibfield  {journal} {\bibinfo  {journal} {Phys. Lett. A}\ }\textbf
  {\bibinfo {volume} {57}},\ \bibinfo {pages} {397} (\bibinfo {year}
  {1976})}\BibitemShut {NoStop}%
\bibitem [{\citenamefont {Abarbanel}, \citenamefont {Brown},\ and\
  \citenamefont {Kennel}(1991)}]{abk1991}%
  \BibitemOpen
  \bibfield  {author} {\bibinfo {author} {\bibfnamefont {H.~D.~I.}\
  \bibnamefont {Abarbanel}}, \bibinfo {author} {\bibfnamefont {R.}~\bibnamefont
  {Brown}}, \ and\ \bibinfo {author} {\bibfnamefont {M.~B.}\ \bibnamefont
  {Kennel}},\ }\bibfield  {title} {\enquote {\bibinfo {title} {Variation of
  {Lyapunov} exponents on a strange attractor},}\ }\href {\doibase
  10.1007/BF01209065} {\bibfield  {journal} {\bibinfo  {journal} {J. Nonlin.
  Sci.}\ }\textbf {\bibinfo {volume} {1}},\ \bibinfo {pages} {175} (\bibinfo
  {year} {1991})}\BibitemShut {NoStop}%
\bibitem [{\citenamefont {Eckhardt}\ and\ \citenamefont {Yao}(1993)}]{eck1993}%
  \BibitemOpen
  \bibfield  {author} {\bibinfo {author} {\bibfnamefont {B.}~\bibnamefont
  {Eckhardt}}\ and\ \bibinfo {author} {\bibfnamefont {D.}~\bibnamefont {Yao}},\
  }\bibfield  {title} {\enquote {\bibinfo {title} {Local {Lyapunov} exponents
  in chaotic systems},}\ }\href {\doibase 10.1016/0167-2789(93)90007-n}
  {\bibfield  {journal} {\bibinfo  {journal} {Physica D}\ }\textbf {\bibinfo
  {volume} {65}},\ \bibinfo {pages} {100} (\bibinfo {year} {1993})}\BibitemShut
  {NoStop}%
\bibitem [{\citenamefont {Doerner}\ \emph {et~al.}(1991)\citenamefont
  {Doerner}, \citenamefont {H{\"{u}}binger}, \citenamefont {Martienssen},
  \citenamefont {Grossmann},\ and\ \citenamefont {Thomae}}]{doe1991}%
  \BibitemOpen
  \bibfield  {author} {\bibinfo {author} {\bibfnamefont {R.}~\bibnamefont
  {Doerner}}, \bibinfo {author} {\bibfnamefont {B.}~\bibnamefont
  {H{\"{u}}binger}}, \bibinfo {author} {\bibfnamefont {W.}~\bibnamefont
  {Martienssen}}, \bibinfo {author} {\bibfnamefont {S.}~\bibnamefont
  {Grossmann}}, \ and\ \bibinfo {author} {\bibfnamefont {S.}~\bibnamefont
  {Thomae}},\ }\bibfield  {title} {\enquote {\bibinfo {title} {Predictability
  portraits for chaotic motions},}\ }\href {\doibase
  10.1016/0960-0779(91)90044-a} {\bibfield  {journal} {\bibinfo  {journal}
  {Chaos Soliton Fract.}\ }\textbf {\bibinfo {volume} {1}},\ \bibinfo {pages}
  {553} (\bibinfo {year} {1991})}\BibitemShut {NoStop}%
\bibitem [{\citenamefont {Johnson}\ \emph {et~al.}(1998)\citenamefont
  {Johnson}, \citenamefont {Mar}, \citenamefont {Carroll},\ and\ \citenamefont
  {Pecora}}]{jmcp98}%
  \BibitemOpen
  \bibfield  {author} {\bibinfo {author} {\bibfnamefont {G.~A.}\ \bibnamefont
  {Johnson}}, \bibinfo {author} {\bibfnamefont {D.~J.}\ \bibnamefont {Mar}},
  \bibinfo {author} {\bibfnamefont {T.~L.}\ \bibnamefont {Carroll}}, \ and\
  \bibinfo {author} {\bibfnamefont {L.~M.}\ \bibnamefont {Pecora}},\ }\bibfield
   {title} {\enquote {\bibinfo {title} {Synchronization and imposed
  bifurcations in the presence of large parameter mismatch},}\ }\href {\doibase
  10.1103/PhysRevLett.80.3956} {\bibfield  {journal} {\bibinfo  {journal}
  {Phys. Rev. Lett.}\ }\textbf {\bibinfo {volume} {80}},\ \bibinfo {pages}
  {3956} (\bibinfo {year} {1998})}\BibitemShut {NoStop}%
\bibitem [{\citenamefont {Boccaletti}\ \emph {et~al.}(2002)\citenamefont
  {Boccaletti}, \citenamefont {Kurths}, \citenamefont {Osipov}, \citenamefont
  {Valladares},\ and\ \citenamefont {Zhou}}]{bocc02}%
  \BibitemOpen
  \bibfield  {author} {\bibinfo {author} {\bibfnamefont {S.}~\bibnamefont
  {Boccaletti}}, \bibinfo {author} {\bibfnamefont {J.}~\bibnamefont {Kurths}},
  \bibinfo {author} {\bibfnamefont {G.}~\bibnamefont {Osipov}}, \bibinfo
  {author} {\bibfnamefont {D.~L.}\ \bibnamefont {Valladares}}, \ and\ \bibinfo
  {author} {\bibfnamefont {C.~S.}\ \bibnamefont {Zhou}},\ }\bibfield  {title}
  {\enquote {\bibinfo {title} {The synchronization of chaotic systems},}\
  }\href {\doibase 10.1016/S0370-1573(02)00137-0} {\bibfield  {journal}
  {\bibinfo  {journal} {Phys. Rep.}\ }\textbf {\bibinfo {volume} {366}},\
  \bibinfo {pages} {1} (\bibinfo {year} {2002})}\BibitemShut {NoStop}%
\bibitem [{\citenamefont {Samanta}\ \emph {et~al.}(2014)\citenamefont
  {Samanta}, \citenamefont {Bhattacharjee}, \citenamefont {Bhattacharyay},\
  and\ \citenamefont {Chakraborty}}]{sam2014}%
  \BibitemOpen
  \bibfield  {author} {\bibinfo {author} {\bibfnamefont {H.~S.}\ \bibnamefont
  {Samanta}}, \bibinfo {author} {\bibfnamefont {J.~K.}\ \bibnamefont
  {Bhattacharjee}}, \bibinfo {author} {\bibfnamefont {A.}~\bibnamefont
  {Bhattacharyay}}, \ and\ \bibinfo {author} {\bibfnamefont {S.}~\bibnamefont
  {Chakraborty}},\ }\bibfield  {title} {\enquote {\bibinfo {title} {On noise
  induced {Poincar{\'{e}}--Andronov--Hopf} bifurcation},}\ }\href {\doibase
  10.1063/1.4900775} {\bibfield  {journal} {\bibinfo  {journal} {Chaos}\
  }\textbf {\bibinfo {volume} {24}},\ \bibinfo {eid} {043122} (\bibinfo {year}
  {2014})}\BibitemShut {NoStop}%
\bibitem [{\citenamefont {Cuomo}\ and\ \citenamefont
  {Oppenheim}(1993)}]{co1993}%
  \BibitemOpen
  \bibfield  {author} {\bibinfo {author} {\bibfnamefont {K.~M.}\ \bibnamefont
  {Cuomo}}\ and\ \bibinfo {author} {\bibfnamefont {A.~V.}\ \bibnamefont
  {Oppenheim}},\ }\bibfield  {title} {\enquote {\bibinfo {title} {Circuit
  implementation of synchronized chaos with applications to communications},}\
  }\href {\doibase 10.1103/PhysRevLett.71.65} {\bibfield  {journal} {\bibinfo
  {journal} {Phys. Rev. Lett.}\ }\textbf {\bibinfo {volume} {71}},\ \bibinfo
  {pages} {65} (\bibinfo {year} {1993})}\BibitemShut {NoStop}%
\bibitem [{\citenamefont {Kocarev}\ and\ \citenamefont
  {Parlitz}(1995)}]{kp1995}%
  \BibitemOpen
  \bibfield  {author} {\bibinfo {author} {\bibfnamefont {L.}~\bibnamefont
  {Kocarev}}\ and\ \bibinfo {author} {\bibfnamefont {U.}~\bibnamefont
  {Parlitz}},\ }\bibfield  {title} {\enquote {\bibinfo {title} {General
  approach for chaotic synchronization with applications to communication},}\
  }\href {\doibase 10.1103/PhysRevLett.74.5028} {\bibfield  {journal} {\bibinfo
   {journal} {Phys. Rev. Lett.}\ }\textbf {\bibinfo {volume} {74}},\ \bibinfo
  {pages} {5028} (\bibinfo {year} {1995})}\BibitemShut {NoStop}%
\bibitem [{\citenamefont {Amritkar}\ and\ \citenamefont {Gupte}(1993)}]{ag93}%
  \BibitemOpen
  \bibfield  {author} {\bibinfo {author} {\bibfnamefont {R.~E.}\ \bibnamefont
  {Amritkar}}\ and\ \bibinfo {author} {\bibfnamefont {N.}~\bibnamefont
  {Gupte}},\ }\bibfield  {title} {\enquote {\bibinfo {title} {Synchronization
  of chaotic orbits: The effect of a finite time step},}\ }\href {\doibase
  10.1103/PhysRevE.47.3889} {\bibfield  {journal} {\bibinfo  {journal} {Phys.
  Rev. E}\ }\textbf {\bibinfo {volume} {47}},\ \bibinfo {pages} {3889}
  (\bibinfo {year} {1993})}\BibitemShut {NoStop}%
\bibitem [{\citenamefont {Junge}\ and\ \citenamefont {Parlitz}(2001)}]{jp01}%
  \BibitemOpen
  \bibfield  {author} {\bibinfo {author} {\bibfnamefont {L.}~\bibnamefont
  {Junge}}\ and\ \bibinfo {author} {\bibfnamefont {U.}~\bibnamefont
  {Parlitz}},\ }\bibfield  {title} {\enquote {\bibinfo {title} {Synchronization
  using dynamic coupling},}\ }\href {\doibase 10.1103/PhysRevE.64.055204}
  {\bibfield  {journal} {\bibinfo  {journal} {Phys. Rev. E}\ }\textbf {\bibinfo
  {volume} {64}},\ \bibinfo {pages} {055204} (\bibinfo {year}
  {2001})}\BibitemShut {NoStop}%
\bibitem [{\citenamefont {Chen}, \citenamefont {Qiu},\ and\ \citenamefont
  {Huang}(2009)}]{cqh09}%
  \BibitemOpen
  \bibfield  {author} {\bibinfo {author} {\bibfnamefont {L.}~\bibnamefont
  {Chen}}, \bibinfo {author} {\bibfnamefont {C.}~\bibnamefont {Qiu}}, \ and\
  \bibinfo {author} {\bibfnamefont {H.~B.}\ \bibnamefont {Huang}},\ }\bibfield
  {title} {\enquote {\bibinfo {title} {Synchronization with on-off coupling:
  Role of time scales in network dynamics},}\ }\href {\doibase
  10.1103/PhysRevE.79.045101} {\bibfield  {journal} {\bibinfo  {journal} {Phys.
  Rev. E}\ }\textbf {\bibinfo {volume} {79}},\ \bibinfo {pages} {045101}
  (\bibinfo {year} {2009})}\BibitemShut {NoStop}%
\bibitem [{\citenamefont {Popovych}, \citenamefont {Tass},\ and\ \citenamefont
  {Hauptmann}(2011)}]{pop2011}%
  \BibitemOpen
  \bibfield  {author} {\bibinfo {author} {\bibfnamefont {O.~V.}\ \bibnamefont
  {Popovych}}, \bibinfo {author} {\bibfnamefont {P.~A.}\ \bibnamefont {Tass}},
  \ and\ \bibinfo {author} {\bibfnamefont {C.}~\bibnamefont {Hauptmann}},\
  }\bibfield  {title} {\enquote {\bibinfo {title} {Desynchronization
  (computational neuroscience)},}\ }\href
  {http://www.scholarpedia.org/article/Desynchronization_(computational_neuroscience)}
  {\bibfield  {journal} {\bibinfo  {journal} {Scholarpedia}\ }\textbf {\bibinfo
  {volume} {6}},\ \bibinfo {pages} {1352} (\bibinfo {year} {2011})}\BibitemShut
  {NoStop}%
\end{thebibliography}
\end{document}